\documentclass[english,aps,prper,reprint,showpacs,titlepage,longbibliography,nofootinbib,superscriptaddress]{revtex4-2}

\usepackage[T1]{fontenc}	
\usepackage[latin9]{inputenc}	
\usepackage{geometry}		
\usepackage{graphicx}
\usepackage[above,below]{placeins}	
\usepackage{times}
\usepackage{multirow}

\usepackage{hyperref}  
\hypersetup{colorlinks=true,urlcolor=blue,citecolor=blue,linkcolor=blue}   
\usepackage{enumitem}        
\setlist{nosep}                 

\usepackage{mathtools}
\usepackage{amssymb}
\usepackage{amsmath}
\usepackage{amsthm}
\usepackage[font=small]{caption}
\usepackage{subcaption}
\usepackage{bbold}
\usepackage{comment}
\usepackage{tabularx}
\usepackage{array}
\usepackage[labelsep=period]{caption}

\makeatletter
\renewcommand{\fnum@figure}{FIG. \thefigure}
\makeatother

\newcommand{\etal}{\textit{et al.}}

\newcommand{\Sin}{$s_{\text{i}}$}
\newcommand{\Sout}{$s_{\text{o}}$}
\newcommand{\Snet}{$s_{\text{n}}$}
\newcommand{\Yin}{$Y_{\text{i}}$}
\newcommand{\Yout}{$Y_{\text{o}}$}
\newcommand{\cCi}{$c^{Ci}$}
\newcommand{\cCo}{$c^{Co}$}
\newcommand{\cHi}{$c^{Hi}$}
\newcommand{\cHo}{$c^{Ho}$}

\begin{document}

\begin{titlepage}

\title{Correlations between student connectivity and academic \\performance: a pandemic follow-up}
\author{Nathan Crossette}
\affiliation{Department of Physics, University of Colorado Boulder, Boulder, Colorado 80309, USA}
\author{Lincoln D. Carr}
\affiliation{Quantum Engineering Program, Colorado School of Mines, Golden, Colorado 80401, USA}
\affiliation{Department of Physics, Colorado School of Mines, Golden, Colorado 80401, USA}
\affiliation{Department of Applied Mathematics and Statistics, Colorado School of Mines, Golden, Colorado 80401, USA}
\author{Bethany R. Wilcox}
\affiliation{Department of Physics, University of Colorado Boulder, Boulder, Colorado 80309, USA}

\begin{abstract}

Social network analysis (SNA) has been gaining traction as a technique for quantitatively studying student collaboration. We analyze networks, constructed from student self-reports of collaboration on homework assignments, in two courses from the University of Colorado Boulder and one course from the Colorado School of Mines. All three courses occurred during the COVID-19 pandemic, which allows for a comparison between the course at the Colorado School of Mines (in a fully remote format) with results from a previous pre-pandemic study of student collaboration at the Colorado School of Mines (in a hybrid format). We compute nodal centrality measures and calculate the correlation between student centrality and performance. Results varied widely between each of the courses studied. The course at the Colorado School of Mines had strong correlations between many centrality measures and performance which matched the patterns seen in the pre-pandemic study. The courses at the University of Colorado Boulder showed weaker correlations, and one course showed nearly no correlations at all between students' connectivity to their classmates and their performance. Taken together, the results from the trio of courses indicate that the context and environment in which the course is situated play a more important role in fostering a correlation between student collaboration and course performance than the format (remote, hybrid, in-person) of the  course. Additionally, we conducted a short study on the effect that missing nodes may have on the correlations calculated from the measured networks. This investigation showed that missing nodes tend to shift correlations towards zero, providing evidence that the statistically significant correlations measured in our networks are not spurious.

\end{abstract}

\maketitle
\end{titlepage}

\section{Introduction}

\subsection{Motivation}

Many studies have demonstrated that interactive engagement, which encourages learning through discussion and collaboration, improves student understanding of physics concepts \cite{Hake1998, Mazur2001}. Additionally, having a sense of belonging within an academic community is associated with persistence and achievement among students, especially among students from underrepresented backgrounds \cite{MotivatingStudents2004, Fischer2007}. Homework assignments in physics courses are one of the primary situations in which students can collaborate and form bonds and a sense of belonging, so studying the relationship between collaboration on homework assignments, connection within a community, and performance may shed light on effective ways to create supportive environments within physics courses.

Social network analysis (SNA) provides a quantitative method for analyzing how individuals can be connected to a larger group. Key concepts related to network analysis are discussed in more detail in Sec.~\ref{Sec:Methods}, but briefly, a social network is a complex network (or graph) with individuals represented by nodes (or vertices), while connections between individuals are represented by links (or edges). In a general network both nodes and links carry weights (indicating characteristics of an individual or the strength of a connection between individuals), and links can have a direction indicating a non-reciprocal type of connection.

\subsection{Background}

SNA is an analytical tool that is gaining traction within the field of physics education research (PER), and there is a small but growing collection of studies examining students' connections to their peers and how it relates to student experiences and outcomes. For example, a pair of studies investigated whether students' connection to their classmates (measured via a series of surveys given periodically throughout the semester) predicts their persistence within the introductory physics sequence~\cite{Zwolak2017, Zwolak2018}. The first study found that centrality (i.e., quantitative measures of a student's level of connectivity to other students; see Sec.~\ref{sec:overview} for mathematical definitions) was a good predictor of persistence. The second study provided a more nuanced analysis that investigated social networks that developed in the physical classroom (the in-class network) and networks that formed outside of the physical classroom (the out-of class network) which incorporated collaboration on homework assignments. This second study found that course grade was more correlated with persistence for students with high final grades (or lack of persistence for students with low final grades), and that centrality in the out-of-class network was more correlated with persistence for ``middle-of-the-pack'' students. With these observations, they concluded that developing social connections outside of the classroom either helped create, or reflected an already existing, commitment to their studies~\cite{Zwolak2018}.

Additionally, SNA has been used to explore whether changes in students' feelings of self-efficacy in physics is related their connection to other students~\cite{Dou2016}. Though this study found that students left their introductory physics courses with a lower average sense of self-efficacy, centrality within the network predicted post-course self-efficacy after controlling for pre-course self-efficacy. Furthermore, centrality measures were associated with various sources of self-efficacy~\cite{Dou2016}. Another study used SNA to analyze student interactions in a help-room setting and determined that the environment was equitable because gender and ethnicity were not predictors of participation~\cite{Brewe2012}. SNA is a subdivision of the larger field of network analysis which uses networks (graphs) to analyze complex systems. Outside of the context of social interactions, networks have been used in PER to study the patterns of student responses to multiple choice surveys ~\cite{Brewe2016, Wells2020, Wheatley2021}.

\subsection{Prior Work}

In a direct precursor to this study, Vargas \etal\ created social networks from students' reports of collaboration on homework assignments in three upper-division courses at the Colorado School of Mines (Mines). Various measures of a student's centrality within the network were then correlated with performance on exams and homework assignments. In all three courses, homework scores were positively correlated with several centrality measures, but negatively correlated with measures representing whether a student collaborated with only a few versus many other students. The findings suggested that students who collaborate both frequently and with many others tended to perform better on graded assignments~\cite{Vargas2018}. Another closely related study examined this same connection between students' performance and their connections to their peers in a highly collaborative introductory physics course and also found a significant link between students' centrality and performance using regression analysis~\cite{Williams2019}.

The study presented in this article builds on the study by Vargas \textit{et al.} to examine the relationship between self-reported student collaboration and performance during the COVID-19 pandemic. Data was collected from one course at Mines and two courses at the University of Colorado Boulder (CU Boulder). This allows for a direct comparison between networks from pre-pandemic and pandemic-affected courses at Mines. Furthermore, the data collection in the CU Boulder courses adds an additional perspective on student collaboration by investigating a different student population and educational context. 

In the following section, Sec.~\ref{Sec:Methods}, we discuss the methodology of the study beginning with a brief overview of key network analytic concepts. Then the context and structure of each of the courses in the study, data collection methods, and our data analysis process are described. Next, in Sec.~\ref{Sec:Results}, we present our results and findings on the relationship between students' collaboration and performance. Then in Sec.~\ref{sec:simulations}, we present a short study of random and simulated networks to provide some perspective on our findings, particularly to address possible impacts of missing data. We end in Sec.~\ref{Sec:Conclusion} with conclusions, discussion of limitations, and future work.

\section{Methodology}
\label{Sec:Methods}

In this section we provide an introduction to relevant network analytic concepts accessible to readers with no prior experience in network analysis. This includes centrality measures, which quantify a student's connection to their peers (Sec.~\ref{sec:overview}); the context of the courses from which data was collected including course format, grading structure, and data collection methods (Sec.~\ref{sec:CUcontext} and Sec.~\ref{sec:MinesContext}); and our data analysis process (Sec.~\ref{Sec:Analysis}).

\subsection{Overview of key network analysis concepts}
\label{sec:overview}
\begin{figure}
  \includegraphics[width=.98\columnwidth]{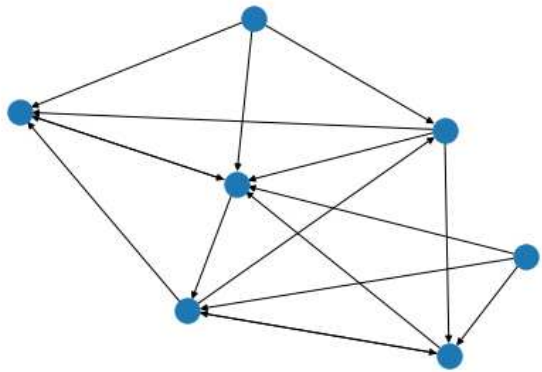}
  \caption{Example of a complex network. Here the blue dots denote nodes (or students) and lines between the notes represent links (or edges). Directionality of the link is denoted by the direction of the arrow on each link.   \label{fig:exampleGraph}}
   \vspace{-1.5\baselineskip}
\end{figure}

An example of a complex network is given in Fig.~\ref{fig:exampleGraph}. In general, links connecting two students have both a weight and a direction. In the networks analyzed in this study, the weight of the link represent the number of times the pair of students reported working together while completing a homework assignment and the direction of the link indicates which student was giving/receiving help, respectively. A network in which the links have a direction is called a \textit{directed network}.

The information about the links in a network is encoded into an \textit{adjacency matrix}. The adjacency matrix, $A$, is an $N\times N$ matrix, where $N$ is the number of nodes, and the matrix elements, $a_{ij}$, are the weights of the links that connect a node $i$ with the node $j$. In a directed network, the adjacency matrix is generally not symmetric since directed connections between individuals are not necessarily reciprocal. Additionally, nodes in a network may also contain information. In our case, this information includes the homework and exam grades of the student represented by the node.

From the networks, we calculate a number of centrality measures, which quantify a node's connection to the rest of the network. Each centrality measure captures a different way in which a node can be connected to the larger network, which in turn can represent different ways in which the node may contribute to the flow of information within the network. The simplest of these centrality measures are the in-strength and out-strength. 

\begin{figure}
  \centering
  \includegraphics[width=.75\columnwidth]{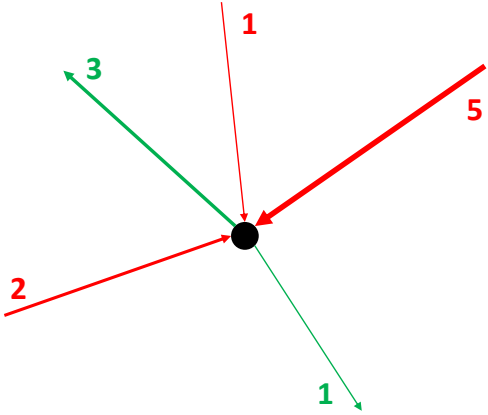}
  \caption{\raggedright  An isolated node with three inward links and two outward links. This node has in-strength $s_{\text{i}}=8$, out-strength $s_{\text{o}}=4$, and net-strength $s_{\text{n}}=4$.   \label{fig:InOutStrength}} \vspace{-1.5\baselineskip}
\end{figure}

In-strength $s_{\text{i}}$ and out-strength $s_{\text{o}}$ quantify the total weight of links terminating and beginning on a node, respectively. The net-strength $s_{\text{n}}$ is simply the difference: $s_{\text{i}} - s_{\text{o}}$. For example, the node in Fig.~\ref{fig:InOutStrength} has an in-strength of eight, an out-strength of four, and a net-strength of four. In our networks, the total number of times a student gave help to other students over the course of the semester is that student's out-strength. The number of times they received help is their in-strength.

The in-disparity $Y_{\text{i}}$ and out-disparity $Y_{\text{o}}$ measure the non-uniformity of a node's inward and outward links, respectively, and provide more information about the distribution of links attached to a node. Nodes with large disparities tend to be connected to very few other nodes within the network, reflecting the large disparity between the few, present connections to other nodes and the many non-existent connections. A node's disparity tends to decrease the more connections the node has to other nodes, and is not defined for nodes without connections. The in- and out-disparities differ from \Sin\ and \Sout. For example, it is possible for a node with a very strong connection to just one other node to have large in- and out- strengths and as well as large disparities. In contrast, a node with many weak connections to other nodes may also have large values for \Sin\ and \Sout, but small values for \Yin\ and \Yout. So, for example, a student who has a strong connection to a few other students will have large disparities and large in- and out-strengths. Alternatively, a student with many relatively weaker connections to other students will have smaller disparities, but could also have large in- and out-strengths depending on the sum of the strengths of the connections. For the equations that describe how the disparity and the in-, out-, and net-strength can be calculated from the adjacency matrix, see the article by Vargas \textit{et al.}~\cite{Vargas2018}.


The closeness $c^C$ and harmonic $c^H$ centralities measure how close a node is to all the other nodes within a network. The closeness centrality of node $i$ is

\begin{equation}
\label{eq:cC_def}
c^C_i = \frac{n-1}{N-1} \frac{n-1}{\sum_{i \neq j} d_{ij}}
\end{equation}

\noindent where $N$ is the total number of nodes in the network, $n$ is the number of nodes reachable from node $i$ (i.e., able to be reached by traversing one-way links), and $d_{ij}$ is the shortest distance between nodes $i$ and $j$. If node $j$ is not reachable from node $i$, then the $d_{ij}$ term is not considered in the sum and the $(n-1)/(N-1)$ pre-factor scales the closeness centrality by the number of reachable nodes. The distance $d_{ij}$ can be related to link weights in a variety of ways (e.i., the distance can some functional dependence on the link weight depending on the context of the network). We discuss our definition of distance in more detail in Sec.~\ref{Sec:Analysis}; however, in our network strong connections mean short distances. The formula above for the closeness centrality was proposed by Wasserman and Faust specifically to account for the case of networks in which some nodes are unreachable from others~\cite{WassermanFaust}.

Another way to address the possibility of certain nodes being unreachable is to use the harmonic centrality which is

\begin{equation}
\label{eq:cH_def}
    c^H_i = \sum_{i \neq j} \frac{1}{d_{ij}}.
\end{equation}

\noindent where $d_{ij}$ is again the shortest distance between node $i$ and node $j$. If a node $j$ is unreachable from node $i$, $d_{ij}$ is effectively infinite and the term in the sum for this pair of nodes is zero. Within the context of social networks, the closeness and harmonic centralities capture the idea of `degrees of separation'. If a student has many connections to their classmates, and if their connections also have many connections (and so on) the first student will have high closeness and harmonic centralities. Additionally, this means that for a student to be unreachable from another, they must share zero mutual collaborators at all levels (i.e., not only do they share no mutual collaborators, their mutual collaborators share no mutual collaborators, and so on).

With regards to directed networks, an important subtlety of the closeness and harmonic centralities is that they can be defined using either the shortest inward directed path or the shortest outward directed path. For example, if a node $n$ has only outward directed links, it is not reachable from other nodes in the network, but other nodes will be reachable from $n$. In this case using the inward shortest paths to compute $c^C$ or $c^H$ will result in centralities of zero, but using the outward paths will result in non-zero centralities. When discussing our results in later sections, we will refer to the closeness and harmonic centralities calculated using the inward shortest paths as $c^{Ci}$ and $c^{Hi}$, respectively, and the quantities calculated using the outward distances as $c^{Co}$ and $c^{Ho}$.

The last centrality measure we will consider is the betweenness centrality. For a node $i$, the betweenness centrality $c^B$ is

\begin{equation}
\label{eq:cB_def}
    c^B_i = \sum_{j,k \in V} \frac{\sigma(j,k|i)}{\sigma(j,k)}
\end{equation}

\noindent where $\sigma(j,k)$ is the number of distinct shortest paths between nodes $j$ and $k$, $\sigma(j,k|i)$ is the number of shortest paths between nodes $j$ and $k$ that pass through node $i$, $V$ is the set of nodes in the network and the sum runs over all pairs of nodes in the network (excluding pairs of nodes containing node $i$)~\cite{Brandes2008}. Conceptually, the betweenness centrality quantifies the extent to which a node is a hub that provides connections between different regions within a network. So, a student who collaborates with two (or more) tight-knit groups which would otherwise be disconnected will have a large betweenness centrality. 

These centrality measures can be broken down into two groups: local centrality measures which only consider a node $i$ and the set of nodes directly connected to $i$, and global centrality measures which depend on the structure of the entire network. The in-strength, out-strength, net-strength, in-disparity, and out-disparity are all local centrality measures while the harmonic, closeness, and betweenness centralities are global measures.

The final network analysis concept relevant to this study (which was not considered in the previous study by Vargas \textit{et al}.~\cite{Vargas2018}) is reciprocity~\cite{Newman_Networks, Cmplx_Nets}. Reciprocity $r$ is only meaningful in directed graphs where the directed links can create an imbalance in the connections between pairs of nodes. For networks with weighted links Squartini \textit{et al}. present the definition of the reciprocity which was used in this study:

\begin{equation}
\label{eq:weightedrecip}
    r = \dfrac{W^{\leftrightarrow}}{W},
\end{equation}

where $W$ is the total weight of all links in the network which can be obtained by summing all the elements of the adjacency matrix~\cite{Squartini2013}. The quantity $W^{\leftrightarrow}$ represents the total weight of reciprocated links. The weight of reciprocal links between a pair of nodes $i$ and $j$ is defined as the minimum of the ``mirrored pair'' (i.e., a matrix element and its partner element in a position reflected across the diagonal) of matrix elements: $w_{ij}^{\leftrightarrow} = \min(w_{ij}, w_{ji})$. With this definition, the total reciprocated weight $W^{\leftrightarrow}$ is

\begin{equation}
    \label{eq:W-leftright}
    W^{\leftrightarrow} = \displaystyle \sum_i \sum_{j \neq i} w_{ij}^{\leftrightarrow}.
\end{equation}

The reciprocity $r$ calculated in Eq.~(\ref{eq:weightedrecip}) represents the reciprocity of the entire network. Conceptually, reciprocity measures the extent to which the connections between nodes are bilateral within the entire network. If a network has few pairs of bilateral links between nodes then the reciprocity will be low, and as links within a network become more bilateral the reciprocity approaches one. In our networks, reciprocity can arise from one of two cases: either from a single student reporting both getting help from and giving help to a second student, or from a pair of students both reporting getting (or giving) help to each other.

\renewcommand{\c}[1]{\multirow{2}{*}{\vspace{1pt}#1}}
\newcolumntype{C}{>{\centering\arraybackslash}X}
\begin{table*}[]
\renewcommand{\arraystretch}{1.1}
  \caption{A summary of the results of the data collection on student collaboration. The final two columns directly compare the level of student collaboration since the courses at Mines and CU Boulder had a different number of homework assignments. \label{tab:StudPart}} \vspace{-\baselineskip}
  \begin{center}
    \begin{tabularx}{\textwidth}{ccCCCccc}
    \hline
    \hline
                        & Participating & Total       & Reports of    &  Reports of   & Sum of edge         & Reports per student &  links per student   \\
                        & students      &  enrollment & getting help  &  giving Help  & weights in network  & per assignment      &  per assignment        \\
      \hline

CU Boulder Fall 2020    & 53            &  83         & 777           &  691          &    1110             &   2.52              &  1.90  \\
CU Boulder Spring 2021  & 48            &  55         & 378           &  335          &     495             &   1.35              &  .94  \\
Mines Fall 2020         & 23            &  27         & 140           &  138          &     208             &   1.73              &  1.29 \\
      \hline
      \hline
    \end{tabularx}
    \end{center}\vspace{-.5\baselineskip}
    
\end{table*}

\subsection{Context}

\subsubsection{CU Boulder: Thermal Physics}
\label{sec:CUcontext}
The two CU Boulder courses from which data was collected occurred during the fall 2020 and spring 2021 semesters. Both courses were an upper-division thermal physics course and taught by the same instructor (BRW) and occurred in a hybrid format with an option to attend synchronous lectures either in-person or remotely and asynchronous lecture recordings available. In both semesters, less than 25\% of the class opted to attend in person. Each course had a total of 12 weekly homework assignments, with collaboration data collected from all but the first assignment.

In the fall 2020 iteration, data on student collaboration was collected through a Qualtrics survey that students completed upon submission of their weekly homework assignments. In the spring 2021 semester, students reported their collaborators directly on their homework solutions. The fall course had three take-home midterm exams and a final with each of the four exams comprising 15\% of a student's final grade. The spring course had only two take-home midterms and a final with each of the three exams comprising 20\% of student's final grade. Students were allowed to submit revisions on both homework assignments and exams for the opportunity to earn back missing points. Students could earn back all missing points on homework assignments but only a fraction of missing points on exams.

Typically, on-sequence physics majors at CU Boulder take this thermal physics course during the fall of their senior year. This is reflected in the total enrollments of the two courses: 83 students took the course in the fall, and 55 took the course in the spring. The process of obtaining consent for collection of student data decreased the number of students from whom data was collected from 83 to 53 in the fall term and 55 to 48 in the spring term. We investigate the possible effects of this missing data in Sec.~\ref{sec:simulation-removal}. CU Boulder is a large, predominantly white research institution with an undergraduate population of roughly 30,000 students with roughly 110 physics majors and 25 engineering physics majors per class year.

\subsubsection{Colorado School of Mines: Math Methods}
\label{sec:MinesContext}
At Mines, collaboration data was collected from an intermediate-level mathematical methods course covering both analytical and numerical methods with significant programming content, during the fall 2020 semester, which was on-sequence with the normal curriculum at Mines. The course was fully remote, synchronous, and taught by author LDC. Students reported their collaborators directly on their homework assignments, just as in the spring iteration of the CU Boulder thermal physics course.

Graded assignments in the course consisted of seven five-point homework assignments, a course project broken into two six 7.5-point assignments across the semester, and 15 points of participation (for a total of 95 possible points). The course had no exams. Mines students also had the opportunity to submit homework revisions to receive points back.

A total of 27 students enrolled in the course, and 23 students consented to the study. Mines is a medium research institution with roughly 5,000 undergraduates, and about 60 physics majors per class year. See Table~\ref{tab:StudPart} for a summary of the course participation.

\subsection{Analysis}
\label{Sec:Analysis}

The first step in processing the collaboration data was to create the adjacency matrix which encodes all students' reports of getting and giving help across the semester. For all three courses, this adjacency matrix was built up assignment by assignment. For each assignment, two separate matrices were created: one containing all reports of getting help (the ``\textit{got-help}'' matrix) for the assignment, and one containing all reports of giving help (the ``\textit{helped}'' matrix) for the assignment. Each of these matrices representing the collaboration on a single assignment contain only ones and zeros. The helped matrix was transposed so that the direction of links would match that of the got-help matrix. The got-help and transposed-helped matrices were then combined using an element-wise logical OR operation, which repeats the analysis performed in the prior study by Vargas \textit{et al.}, to create one ``combined'' matrix for each homework assignment~\cite{Vargas2018}. Finally, the individual combined matrices for each assignment were summed element-wise to create the ``total'' adjacency matrix $A_{\text{tot}}$ representing all collaboration during the course. As a final step, the diagonal of the adjacency matrix was set to zero to ignore any cases of a student reporting themself.

This total adjacency matrix can be used to directly compute \Sin, \Sout, \Snet, \Yin, and \Yout\ for each node according to equations which can be found in~\cite{Vargas2018}. To compute the closeness, harmonic, and betweenness centralities, the matrix of distances $d_{ij}$ must first be calculated. Entries in $A_{\text{tot}}$ range from zero to eleven in the CU Boulder courses and from zero to seven in the Mines course, and large values in $A_{\text{tot}}$ represent a large amount of collaboration between students. Large values in $A_{\text{tot}}$ should correspond to \textit{short} distances in $d_{ij}$. To create a $d_{ij}$ consistent with this relationship between collaboration and distance, we took the reciprocal of the elements of $A_{\text{tot}}$, unless the element was zero in which case it remained zero. 

The calculation of the closeness, harmonic, and betweenness centralities was done using built-in functions from the NetworkX python package. When calculating the closeness and harmonic centralities in a directed network, the NetworkX functions default to calculating \cCi\ and \cHi. To calculate \cCo\ and \cHo, the directions of all links in the adjacency matrix are swapped which is accomplished, in practice, by transposing the adjacency matrix.

The output of the functions that calculate the centrality measures are dictionaries with the node label as the key and the centrality measure of the node as the value. With the dictionaries, the Pearson correlations between the centrality measures and students' homework and exam grades were directly calculated. The students in the CU Boulder courses could submit revisions to exams to receive points back, but we calculated correlations using the pre-revised exam scores. Correlations with homework scores were calculated using the post-revised grade as the pre-revised grade was never recorded for these courses.

Only reports from students consenting to the study were included in the construction of the social network. In some instances, though, a student would report collaboration with a student who had not consented to the study. In this case, a node and appropriate links (depending on the reports from consenting students) for the non-consenting student would be added and included in the final network representing only information reported from the other, consenting student (i.e., collaboration reports submitted by non-consenting students as part of the normal coursework were not included in the data collection or the network). Thus, some non-consenting students appear in the final networks, the presence of which contributes to the centrality measures for other students in the network. However, the non-consenting students were not included in the calculations of correlations as data on their course scores was not collected as part of the study.

\section{Results}
\label{Sec:Results}

\begin{figure}[h!]

  \centering
    \begin{subfigure}{.975\columnwidth}
    \centering
    \includegraphics[width=\columnwidth]{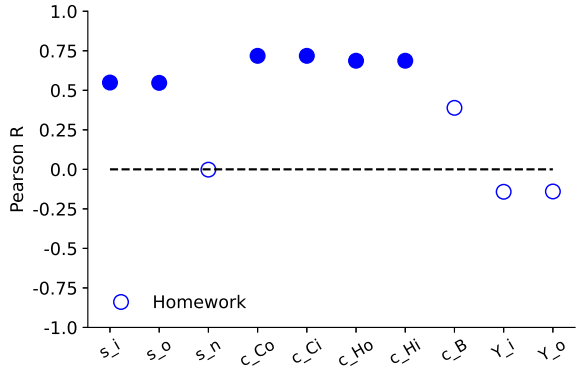}
    \caption{Fall 2020 Math Methods at Mines. \label{fig:Mines_Corrs}}
  \end{subfigure}
  \begin{subfigure}{.975\columnwidth}
    \centering
    \includegraphics[width=\columnwidth]{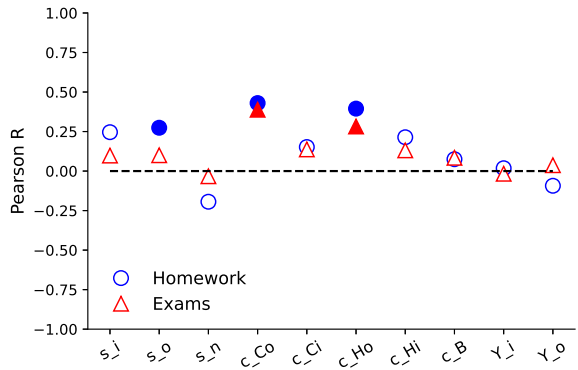}
    \caption{Fall 2020 Thermal Physics at CU Boulder. \label{fig:CUFA_corrs}}
  \end{subfigure}
  \begin{subfigure}{.975\columnwidth}
    \centering
    \includegraphics[width=\columnwidth]{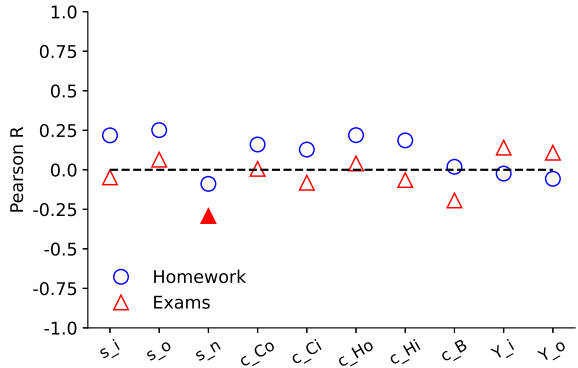}
    \caption{Spring 2021 Thermal Physics at CU Boulder. \label{fig:CUSP_Corrs}}
  \end{subfigure}

  \caption{Correlations between nodal centrality measures and student performance in three pandemic-affected courses. Correlations significant at the $p<.05$ level are indicated with filled markers. The math methods course at Mines had no exams. \label{fig:corrResults}} \vspace{-1.5\baselineskip}
\end{figure}

One of the main goals of this study was to compare the networks and correlations between student centrality and performance between courses taking place before the COVID-19 pandemic and courses occurring during the pandemic. An overview of the collected data in this study is provided in Table~\ref{tab:StudPart}. In all three courses, students more often reported getting help than giving help, though the difference was smallest by far in the course at Mines. This observation cannot be compared to the prior study since only aggregated adjacency matrices remain from the prior study, but it suggests there may be a different conceptualization about the nature of giving and receiving help between students at Mines and CU Boulder.

Likewise, the frequency of reporting also cannot be directly compared between the pre-pandemic and pandemic-affected courses, but the density of the networks, in terms of the number of links per student per assignment present in the full networks can be compared. In the courses analyzed in the prior study, the classical mechanics, electromagnetism, and quantum mechanics courses had 1.4, 1.5, and 1.7 links per student per assignment. Except for the spring 2021 thermal physics course at CU Boulder, the level of collaboration occurring between students in the pandemic-affected courses is relatively similar to the pre-pandemic courses (see Table~\ref{tab:StudPart}). In the fall 2020 course at CU Boulder and the course at Mines, students, on average, gave or received help from more than one other student on each assignment (see Table~\ref{tab:StudPart}). This may be unexpected since the COVID-19 pandemic made face-to-face collaboration more difficult between students; however, it is also possible that students' definition and threshold for reporting giving and receiving help may have changed due to the pandemic and online modes of interaction became more common.

The correlations between centrality measures and student performance in the three pandemic-affected courses are shown in Fig.~\ref{fig:corrResults}. The results from the math methods course at Mines very closely matched the patterns seen in the pre-pandemic courses at Mines~\cite{Vargas2018}. The fall 2020 course at Mines had strong correlations between performance and the in-strength, out-strength, closeness centrality, and harmonic centrality. Furthermore, the non-significant correlations between the net-strength and betweenness centrality also matched the pre-pandemic findings. The largest difference between the pre-pandemic and pandemic-affected courses at Mines is that during the COVID-19 pandemic, the in-disparity and out-disparity were not correlated statistically significantly with performance (as they were in the prior study). Among the three pre-pandemic courses, though, the in- and out-disparities correlated negatively with performance in two of the courses, but not the third (Quantum Mechanics). The negative correlation suggested that students who collaborated with many other students (as opposed to only few other students) tended to get higher scores~\cite{Vargas2018}. This replication of the patterns from the pre-pandemic courses indicates that the environment at Mines was largely able to preserve the connection between collaboration and performance during the pandemic despite the course being fully remote. 

The fall 2020 iteration of the thermal physics course at CU Boulder showed some of the same patterns as the courses from Mines. The outward closeness centrality and outward harmonic centrality were positively correlated with performance on homework assignments and exams at the $p<.05$ level with performance on homework assignments, and the out-strength (in addition to \cCo and \cHo) was positively correlated with performance on homework assignments at the $p < .05$ level. Since outward directed links from a node represent giving help to other students, the statistical significance of only the outward oriented centrality measures indicates that students who were a source of help close to many other students in the course tended to get higher grades. This could suggest that students doing well in the course were more able to provide help to others, or that providing help to others (but not receiving help) improves a student's performance.

The goal of this analysis was to compare patterns of correlation in these courses to the pre-pandemic data; thus the analysis does not focus on the significance of individual correlations, but rather the overall pattern of the correlations. However, multiple statistical tests were performed with this data, introducing a risk of false positives (a type 1 error). To address this possibility, we applied the conservative Bonferroni test which effectively lowers the p-value threshold required to reject a null hypothesis at the $p < .05$ level. The test divides the initial p-value threshold (of .05) by the number of statistical tests performed which in this case was ten tests per course for a stricter threshold of $p < .005$. Table~\ref{tab:Bonferroni} shows the results of the application of this stricter criterion on the significance levels of the correlations.

\begin{table}
\renewcommand{\arraystretch}{1.1}
  \caption{A summary of the application of the Bonferroni test to the significance levels of the measured correlations between student centrality (listed in the first column) and homework scores. P-values for correlations that satisfy the Bonferroni test are indicated in bold with asterisks, meaning that these correlations are especially strong and resistant to a false positive analysis. \label{tab:Bonferroni}}\vspace{-\baselineskip}
  \begin{center}
    \begin{tabularx}{\columnwidth}{Ccc}
    \hline
    \hline
    Correlation  & Correlation Value & p-value  \\
      \hline
    Mines \Sin       & .549  & .0067  \\
    Mines \Sout      & .547  & .0069  \\
    Mines \cCo       & .718  & \textbf{.00011*} \\
    Mines \cCi       & .718  & \textbf{.00011*} \\
    Mines \cHo       & .687  & \textbf{.00029*} \\
    Mines \cHi       & .687  & \textbf{.00029*} \\
    CU Fall \Sout    & .274  & .047   \\
    CU Fall \cCo     & .430  & \textbf{.0013*}  \\
    CU Fall \cHo     & .395  & \textbf{.0034*}  \\
    CU Spring \Snet  & -.292 & .044   \\
      \hline
      \hline
    \end{tabularx}
    \end{center}\vspace{-.5\baselineskip}
    
\end{table}

The nuances of this distinction between the inward closeness/harmonic centralities and the outward closeness/harmonic centralities were not explored in the prior study. Both before and during the pandemic, in the courses at Mines the correlations for the inward and outward centralities were nearly identical rendering a discussion of the distinction somewhat irrelevant. In both courses at CU Boulder, however, different correlations were observed which indicated that being a well-connected source of help was associated with high performance.

To investigate the different correlations between the inward-directed versus outward-directed closeness/harmonic centralities and grades observed in the CU Boulder courses (and lack of difference seen in the Mines courses, see Fig.~\ref{fig:corrResults}) we looked at the reciprocity of each network. Reciprocity measures the degree to which a connection between nodes is bi-directional (i.e., that an edge from node $i$ to node $j$ has a matching edge from $j$ to $i$). Using the method proposed by Squartini \textit{et. al.} for calculating reciprocity in weighted networks~\cite{Squartini2013}, we found that all networks were highly reciprocal.

\begin{table}[b]
\renewcommand{\arraystretch}{1.1}
  \caption{The reciprocity for each of the courses in the present study. The reciprocity takes on values from 0 (for no reciprocity) to 1 (completely reciprocal). The method proposed by Squartini \textit{et. al} was used to calculated the reciprocity taking edge weight into account~\cite{Squartini2013}. \label{tab:recip}} \vspace{-\baselineskip}
  \begin{center}
    \begin{tabularx}{\columnwidth}{lC}
    \hline 
    \hline
    Course & Weighted Reciprocity \\  
    \hline
    Mines Fall 2020         & .990 \\ 
    CU Boulder Fall 2020    & .861 \\ 
    CU Boulder Spring 2021  & .848 \\ 
    \hline
    \hline
    \end{tabularx}
  \end{center}\vspace{-1.5\baselineskip}
\end{table}

The reciprocities calculated in Table~\ref{tab:recip} are higher than typically seen in social networks~\cite{Squartini2013, Newman_Networks}. In the case of the Mines course, this finding appears consistent with the symmetry in the correlations between inward- versus outward-directed closeness and harmonic centralities and grades. Because, if for every outward directed edge there is a corresponding (equally weighted) inward edge, then the outward shortest distances from a node to all other nodes will be identical to the inward shortest distances. The reciprocities in the CU Boulder networks were lower than those for the Mines network, though still relatively large. Despite these large reciprocities, there is an asymmetry between the correlations with the inward versus outward centrality measures.

Such a result may seem counter intuitive, but it is not unexpected since reciprocity within a directed network, in general, is not related to the symmetry of its adjacency matrix~\cite{Squartini2013, Garlaschelli2010}. The larger asymmetry between students' reports of receiving versus giving help is a possible explanation for the asymmetry in the correlations, but is not, by itself, sufficient to account for the asymmetry since the networks simulated in Sec.~\ref{sec:simulations} were constructed with an asymmetry in reporting, but did not reproduce the asymmetry in the correlations. One possible explanation for this lack of consistency in the correlations at Mines and CU Boulder is a cultural difference in how students define thresholds for giving and receiving help or in how students collaborate at the two institutions.

Correlations between a student's centrality and performance were almost non-existent in the spring 2021 iteration of the thermal physics course at CU Boulder. The only statistically significant correlation was between the students' net-strength and exam performance. This correlation was negative, and since the net-strength is the in-strength minus the out-strength, indicating that students who received more help than they gave tended to get lower scores on exams. This result appears consistent with and somewhat complementary to the results from the fall iteration of the course that students who provided help tended to perform better. Another feature of this course from the spring of 2021 was that it had the least level of student interaction in terms of links and reports per student (see Table~\ref{tab:StudPart}) and was the only course which contained multiple, disconnected components\footnote{Individual components within a network are subsets of connected nodes within a network that connected to each other, but completely disconnected from nodes in other components. The network in Fig.~\ref{fig:exampleGraph} has only one component.}. This could be a consequence of the course occurring in the off-sequence semester, thus students in the spring semester course may be less likely to know each other and less likely to have taken prior courses together. Alternatively, students may have been less engaged overall in the spring due to increased pandemic-related burnout, an effect which may be compounded by the large proportion of second semester seniors in the course who were approaching graduation. All these possible explanations are speculation; interviews with students in the course would be necessary to provide more insight into these results.

Overall, these results show that the format of the course (whether in-person, remote, or hybrid), does not appear to dictate whether there will be a correlation between student collaboration and performance. The environment at Mines maintained the connection despite a fully remote instruction format during the COVID-19 pandemic, while the hybrid instruction format at CU Boulder was not able to consistently produce a connection despite having an option for in-person lectures during both semesters. To determine whether the correlation is just typically weaker at CU Boulder or whether the variation in results between the spring and fall semesters is due to on/off-sequence effects versus pandemic related burnout more research would need to be done after a return to normal instruction.



\section{Statistical simulations of social networks}
\label{sec:simulations}

Networks are complex, non-linear objects. A small perturbation in a network could have anywhere from a negligible to a large affect on the network depending on the location and nature of the perturbation. As in any experimental study, our data collection is susceptible to random and statistical errors. In particular, a cursory analysis of the consistency of student reporting in this study indicates that students may have different thresholds for what qualifies as giving or receiving help which can result in either missing or spurious links. The effect of this inconsistency is partially addressed by combining student reports using a logical OR operation as discussed in Sec.~\ref{Sec:Analysis}, but as with any human subjects research, some level of human error is expected. Additionally, collaboration reports from students who did not consent to the study were not included in the construction of the social networks resulting in both missing links and missing nodes.

To better understand the significance of our findings given the limitations of the data collection process, we conducted analyses of random networks and the effect that removing nodes from networks has on the correlations between centrality measures and performance. Due to the rather large proportion of missing nodes in the fall 2020 course at CU Boulder, most of our analyses focus particularly on understanding how the results from this semester's course may have been affected by missing nodes. As noted by a recent study on the effects of missing data on robustness of centrality measures, most applied network studies do little more than acknowledge that measurement errors may have occurred \cite{martin_niemeyer_2019}; so, in addition to better understanding the significance of our results, we hope to expand the knowledge of, and introduce a practice of, error analysis into social network research within the PER field. 

\begin{figure}
  \centering
  \includegraphics[width=.9725\columnwidth]{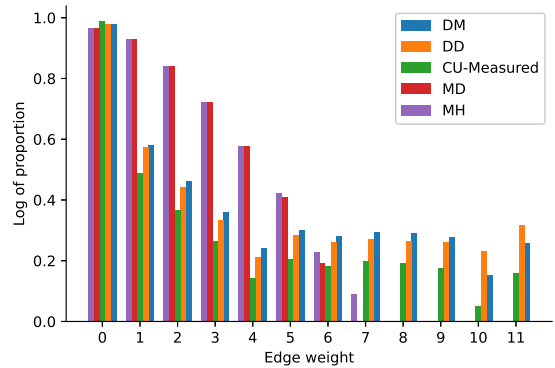}
  \caption{\raggedright Distribution of edge weights in the fall 2020 CU Boulder thermal physics course and in simulated networks. The height of the bars is the log of the ratio of the number of links of a particular weight to square of the number of nodes, i.e., the total number of ordered pairs of nodes. \label{fig:edgeDistrib}} \vspace{-1.5\baselineskip}
\end{figure}

\subsection{Simulated courses using random networks}
\label{sec:simulation-randnets}

To provide a deeper perspective, beyond the simple application of a t-test, on the significance of our results, we developed four methods for constructing random networks. Real world networks, generally, are poorly modeled by random networks~\cite{NewmanPhysicsToday}, so we attempted to create models which more closely matched the structure of the networks we measured. Simulating networks directly explores randomness in networks and helps to demonstrate how our results compare to ensembles of networks with similar characteristics, and demonstrate that the results we reported in Sec.~\ref{Sec:Results} are unlikely to be due to random chance. 

Data used to simulate the networks were taken from the fall 2020 CU Boulder thermal physics course as it was the course with the largest enrollment but the smallest fraction of participating students. While accumulating the network for this course from the 11 homework assignments, all reports of getting help were compiled into a ``got-help'' adjacency matrix. A second matrix containing all the reports of helping another student, the ``helped'' matrix, was also compiled. Elements in these matrices had values between 0 and 11 (inclusive), and the edge distribution represents the relative abundance of each of the possible edge strengths between nodes.


The most basic method for simulating social networks, which will be referred to as the Degree Match (DM) method, used the edge distributions from the CU Boulder fall 2020 course's got-help and helped matrices to construct simulated got-help and helped matrices which matched the distribution of edge weights seen in the fall 2020 CU Boulder thermal physics course. Then the simulated helped matrix was transposed for the sake of directly matching the analysis performed on the real networks. The two simulated adjacency matrices were then combined using an element-wise maximum function to mimic the effect of using the logical OR operation in the analysis of the real network data. As expected, the DM method succeeded in matching the distribution of edge weights seen in the measured networks (see Fig.~\ref{fig:edgeDistrib}).

The second method, which will be called the Multiple Homeworks (MH) method, uses the simple probability that a pair of students collaborated to create 11 pairs of got-helped and helped adjacency matrices, one pair for each homework assignment. Unlike the DM method, the matrices simulated in the MH method only contain ones and zeros which were pulled from a Bernoulli distribution. The helped matrices in each pair was transposed to match the original analysis process, then a logical OR operation was applied assignment-by-assignment and the total network is created by accumulating the combined helped/got-help adjacency matrices over all assignments. In further dissimilarity with the DMs method, the MH method does not succeed in matching the distribution of edge weights seen in the real networks. As can be seen in Fig.~\ref{fig:edgeDistrib}, the MH method results in more low-weight links and fewer high-weight links between nodes that what was seen in the fall 2020 course at CU Boulder.

\begin{figure}
  \centering
  \includegraphics[width=.9725\columnwidth]{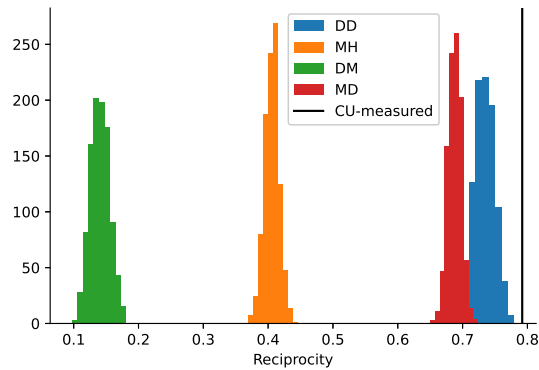}
  \caption{\raggedright Distributions of reciprocities for each of the four random network methods developed in this study. For each method, 1000 random networks were created. \label{fig:simRecipHist}} \vspace{-1.5\baselineskip}
\end{figure}

In both of these methods above, the helped and got-help adjacency matrices were created from independent probability distributions. However, as noted in Sec.~\ref{Sec:Results}, our measured networks were all highly reciprocal. Upon analysis of the random networks generated with the MH and DM methods, we found they lacked the level of reciprocity seen in the real networks (see Fig.~\ref{fig:simRecipHist}). This motivated the development of our third and fourth methods to make our simulations better match the real networks.

The Multiple Dependent (MD) method is identical to the MH method except it only randomly generates 11 got-help adjacency matrices (with the same probability of an edge as the MH method) to simulate the 11 homework assignments. Then, the helped adjacency matrices for each assignment are generated depending on the respective got-help adjacency matrix. If there was a got-help report between a pair of students (i.e., student A reports getting help from student B) there was a relatively large probability that there would also be a matching report of helping between the students in the helped matrix (specifically, student A would also report giving help to student B). If there was no report of getting help between a pair of students, there was still a small probability that a report of helping would exist. While this method better matched the the level of reciprocity in the real network (see Fig.~\ref{fig:simRecipHist}), the MD method fails to match the edge distribution seen in the real networks, just as with the MH method (see Fig.~\ref{fig:edgeDistrib}).

Our final method, called the Direct Dependent (DD) method, comes the closest to matching the level of reciprocity seen in the real networks (see Fig.~\ref{fig:edgeDistrib}). The DD method creates a single got-help adjacency matrix representing reports of getting help from across all homework assignments using the same edge distribution as the DM method. However, similar to the MD method, the adjacency matrix representing all reports of giving help across the course is created based on the got-help matrix. So, if there was a report of getting help between a pair of students (say student A reports getting help from student B), there was a high likelihood that there would be a reciprocal report of giving help between the pair (i.e., student A would also have given help to student B). Similar to the DM method, the DD method generated networks that better matched the edge distribution of the real networks. A summary of the key features of the methods are supplied in Tab.~\ref{tab:SimMethods}.

These second two methods (MD and DD) required some tuning of the dependent probabilities, but we were successful in creating networks with the correct number of total students reports of getting and giving help we saw in the real networks. Furthermore, both methods greatly increased the reciprocity we saw in the simulated networks, nearly to the level seen in the real networks (see Fig.~\ref{fig:simRecipHist}). Importantly, this finding indicates that the reciprocity seen in the real networks is not a consequence of combining student reports using a logical OR operation, but rather a real signal that student collaboration tends to be highly reciprocal since low reciprocity was seen in the DM and MH networks despite the use of the logical OR combination method.



\begin{table}[b]
\renewcommand{\arraystretch}{1.1}
  \caption{A summary of the key features of the four methods developed to simulate networks of student collaboration. The row for independent/dependent sub-networks refers to whether the got-help and helped matrices were created independently or whether the helped matrix depended on the got-help matrix. \label{tab:SimMethods}} \vspace{-\baselineskip}
  \begin{center}
    \begin{tabularx}{\columnwidth}{lCC}
    \hline 
    \hline
     & Simulates multiple assignments & Single matrix for all assignments \\ 
    \hline
    Independent    & Multiple HWs       & Degree Match      \\
    sub-networks   & (MH)               & (DM)              \\
    Dependent      & Multiple Dependent & Direct Dependent  \\
    sub-networks   & (MD)               & (DD)              \\  
    \hline
    \hline
    \end{tabularx}
  \end{center}\vspace{-1.5\baselineskip}
\end{table}

\begin{figure}

  \centering
  \begin{subfigure}{.975\columnwidth}
    \centering
    \includegraphics[width=\columnwidth]{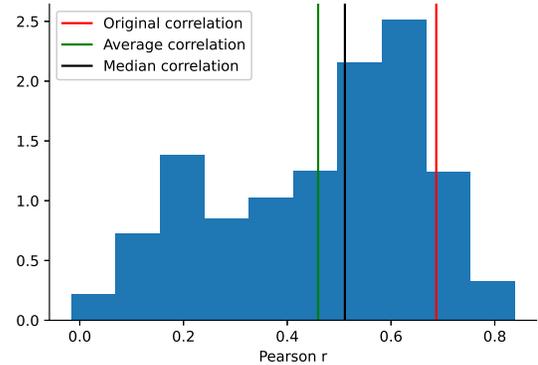}
    \caption{Histogram of correlations in reduced networks created by preferentially dropping nodes with low grades from the Mines network. \label{fig:LowGradeMinesDrop}}
  \end{subfigure}
  \begin{subfigure}{.975\columnwidth}
    \centering
    \includegraphics[width=\columnwidth]{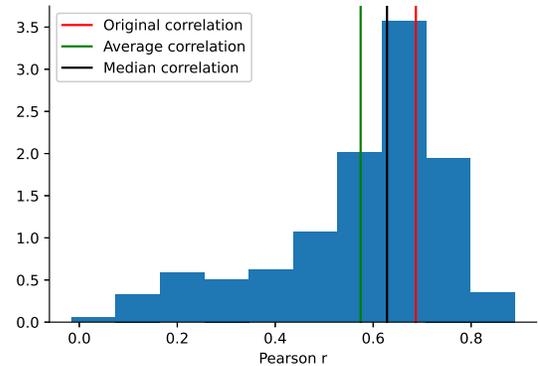}
    \caption{Histogram of correlations in reduced networks created by preferentially dropping nodes with low centrality from the Mines network. \label{fig:LowCentMinesDrop}}
  \end{subfigure}

  \caption{Histogram of correlations between grades and inward harmonic centrality, \cHi, in networks created by dropping nodes from the full Mines network. The red lines in each plot represent the correlation between grades and inward harmonic centrality measured in the original Mines network. The green and black lines represent the mean and median correlation of the ensemble of reduced networks. The blue bars represent the histogram of correlations calculated from the networks in the ensemble. \label{fig:SimulatedMinesDrops}} \vspace{-1.5\baselineskip}
\end{figure}

\subsection{Simulated removal of nodes}
\label{sec:simulation-removal}

To investigate the effect of missing nodes within our networks, we simulated the removal of nodes from the Mines network and from networks generated using the DD method described in the previous section. This method was chosen since it roughly matched the distribution of links seen in the real network, and came closest to matching the reciprocity of the real network. Nodes were missing from all networks, but were most prevalent in the fall 2020 course at CU Boulder which was missing nodes for 36\% of the course's total enrolled population. In this analysis, we consider there to be a `true,' `complete' network which accurately and precisely represents student collaboration in a course. From the `true' network we will remove nodes to make `reduced' networks, then see how the correlation between centrality and grades is affected in the reduced networks. 

Five different methods for dropping students from the networks were tested. Each method used a different probability distribution to select students to drop from the network. Two probability distributions determining a node's likelihood to be dropped were created based on grades: one where high-performing students had a large probability of being dropped and another which gave low-performing students a higher likelihood of being removed. Two more distributions were created based on centrality: one in which students with high centrality were more likely to be dropped, and a second in which students with low centrality were more likely to be dropped. The final method removed students randomly, i.e., all students had the same probability of being removed.

Dropping nodes from the Mines network provides insight into the effect removing nodes from a `complete' network which has a strong correlation between centrality and grades. For each of the five methods described above, 1000 sets of eight distinct nodes were selected for dropping from the network to match the proportion of missing nodes from the fall semester course at CU Boulder. The selected students were dropped to create reduced networks, and the centralities for the remaining nodes were recalculated. The new centralities were then correlated with the grades of the students remaining in the reduced network. The resulting distribution of correlations in the 1000 reduced networks are shown in Fig.~\ref{fig:SimulatedMinesDrops} for the methods preferentially dropping students with low grades and low centrality. We chose to show these plots since we suspected that students with low grades or low centrality were less likely to participate in the study, report their collaborators, and more likely to be disconnected from the larger network. The results in Fig.~\ref{fig:SimulatedMinesDrops} represent the pattern seen in all the five dropping methods since the all methods gave qualitatively similar results.

As can be seen in the histograms in Fig.~\ref{fig:SimulatedMinesDrops}, very few reduced networks had a correlation larger than the correlation measured in the complete network (i.e., there are relatively fewer counts above the red lines in the histograms). Furthermore, the median correlation splits the distribution of correlations in half, so when removing nodes from the Mines network there is an equal probability of getting a correlation above or below the median. The median correlation among the reduced networks falls below the original correlation, indicating that removing nodes from a complete network with a strong correlation between grades and centrality tends to decrease the correlation measured in a reduced network. Removing nodes using the other three methods (high grades, high centrality, and randomly) all produced similar results as seen in Fig.~\ref{fig:SimulatedMinesDrops}.

To extend the analysis of the effect of missing nodes, a similar dropping process was performed on random networks generated using the DD method described in the previous section. To create the probability distributions for dropping nodes by grades, the homework scores from the fall 2020 CU Boulder thermal physics course were applied randomly to the nodes of the simulated network. This process of randomly assigning grades to nodes resulted in complete networks that typically did not have significant correlations between grades and centrality (specifically, the distribution of correlations closely fit to a t-distribution centered at zero, as would be naively expected). So, while dropping nodes from the Mines network helped to show what happens when nodes are removed from networks with statistically significant correlations between grades and centrality, removing nodes from `complete' networks without significant correlations helps show the likelihood of us finding a spurious correlation in the `reduced' network. 

\begin{figure}

  \centering
  \begin{subfigure}{.975\columnwidth}
    \centering
    \includegraphics[width=\columnwidth]{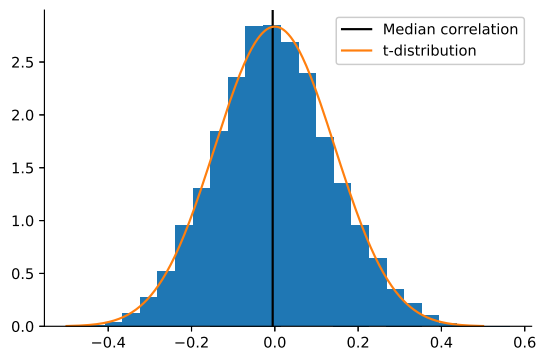}
    \caption{Preferentially dropping nodes with high centrality. \label{fig:HighCentDrop}}
  \end{subfigure}
  \begin{subfigure}{.975\columnwidth}
    \centering
    \includegraphics[width=\columnwidth]{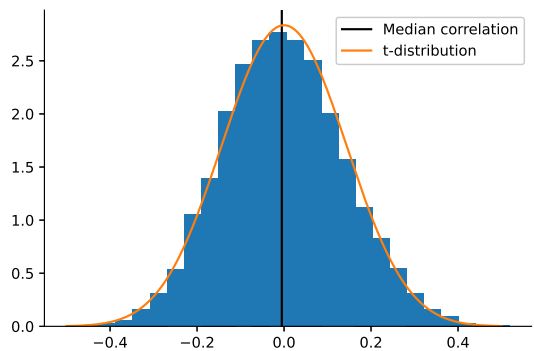}
    \caption{Preferentially dropping nodes with low centrality. \label{fig:LowCentDrop}}
  \end{subfigure}

  \caption{Histogram of correlations between grades and inward harmonic centrality, \cHi, in networks simulated with the DD method and 36\% of nodes removed based on the nodes' centrality. \label{fig:RandomDrops}} \vspace{-1.5\baselineskip}
\end{figure}

Two approaches were taken to study the effect of missing nodes from these simulated networks. First, a single network was constructed (with 83 nodes), then 30 nodes were dropped, and correlations re-calculated to directly mimic the loss of nodes from the fall 2020 course at CU Boulder. This approach was applied iteratively and allowed for a large number of simulated networks to be analyzed. The distribution of resulting correlations in the reduced networks (after dropping the most and least central student) are shown in Fig.~\ref{fig:RandomDrops}. The average and original correlations are not shown (as in Fig.~\ref{fig:SimulatedMinesDrops}) since they were all nearly equal to zero. The distribution of correlations in the reduced networks also fit very well to a t-distribution centered on zero, suggesting that dropping students does not tend to produce a net shift the correlation measured in the reduced networks.

The second approach more closely matched the process applied to the Mines network. In this approach, a single network (of 83 nodes) was constructed, but instead of choosing only one set of nodes to drop, 500 sets of 30 nodes were produced and each set was dropped from the complete network to create 500 different reduced networks. New correlations were then calculated in all of the reduced networks before repeating the process with a new simulated complete network. This process repeated for 500 random `complete' networks. This method allowed for a more detailed perspective on dropping nodes, but fewer networks could be analyzed with this process. All five methods for selecting nodes to drop were applied in this analysis and in each case the result was similar: dropping nodes tended to shift the correlation measured in the reduced network towards zero, and the proportion of complete networks with a non-significant correlation that became reduced networks with a significant correlation when dropping nodes was less than three percent.

When taken together, the results from dropping nodes from the Mines and simulated networks suggests that the statistically significant correlations between homework scores and network centrality measured in the fall 2020 thermal physics course at CU Boulder likely are not spurious results caused by missing data.

\vspace{-.5\baselineskip}

\section{Summary and Conclusion}
\label{Sec:Conclusion}
We collected data on student collaboration in two physics courses at the University of Colorado Boulder and once course at the Colorado School of Mines. The courses occurred in the midst of the COVID-19 pandemic which allowed for a partial comparison to the results of a prior study which occurred before the implementation of remote and hybrid courses as a result of the pandemic. Social networks were constructed based on students' reports of giving and receiving help on homework assignments throughout each course. From the networks we calculated nodal centrality measures which quantify the level of connection a student has to the rest of their classmates.

When calculating the correlations between students' centrality and their performance on homework assignments and exams we found different results in each of the courses. The results from intermediate math methods course at Mines closely matched the results from the pre-pandemic study: there were statistically significant correlations between students' in-strength, out-strength, closeness centrality, and harmonic centrality which indicates that students who collaborate frequently and are closely connected to their peers tend to get higher grades on homework assignments.

The courses at CU Boulder had less connection between students' centrality and course performance. Despite having a higher density of links than the math methods course at Mines, the fall 2020 thermal physics course at CU Boulder only had statistically significant correlations between homework grade at the out-strength and the harmonic and closeness centralities calculated using the out-ward directed shortest paths from a node (exam scores were only correlated with the latter two centralities). This indicates that students who tend to provide help to others, and are close sources of help to their classmates tended to score better on the course's homework assignments and exams. In the spring 2021 course at CU Boulder, there was only a negative correlation between exam score and net-strength (representing the net amount of help given or received). This suggests that students receiving more help than they gave tended to score lower on this course's exams.

The consistency between the results found in this study and the pre-pandemic results from Vargas \textit{et al.} show that the environment at Mines was able to preserve a connection between collaboration and course performance despite a fully remote course format. When contrasted with the lack of consistency in the results from hybrid courses at CU Boulder, this study demonstrates that course context is important for creating a connection between student collaboration and performance. For example, the on-sequence iteration of the thermal physics course at CU Boulder did show some correlation between collaboration and performance, while the off-sequence iteration showed nearly no correlation.

One of the primary limitations to this study stems from limitations in the data collection process. This was specifically a concern for the fall 2020 course at CU Boulder which was missing roughly 36\% of nodes representing student who did not consent to the data collection process. To address possible effects caused by missing data, we constructed several models for generating networks which resulted in one which was able to match the edge distribution and reciprocity seen in the measured networks. The simulated networks assumed that the level of collaboration in the missing part of the network matched the level seen in the measured networks. Simulations of random networks with this method corroborated the results of the t-test which established the statistical significance of our findings.

The development of the network simulation models also allowed for an investigation of the effect that removing nodes from larger networks has on the correlations between centrality and course performance. `Complete' networks of 83 nodes were created and sets of 30 nodes were selected for removal by various metrics. The removal of nodes tended to shift correlations towards zero and resulted in a non-significant correlation (in the complete network) to shift to a significant correlation (in the reduced network) in less than three percent of trials. A similar approach to dropping nodes was applied to the measured network from the math methods course at Mines, which produced similar results: a tendency to decrease correlations. These results suggest that the statistically significant correlations measured in the fall 2020 course at CU Boulder reflect statistically significant correlations in the complete network.

Though prior research in education shows that collaborative interaction among students generally leads to better learning outcomes (at least in part due to developing a student's sense of belonging), lacking a correlation between student performance and centrality should not be considered a necessarily undesirable feature. In the case of commuting students, students who are working while attending school, recently transferred students, or other situation in which students are less able to collaborate with their peers, an ideal course would overcome the obstacles faced by these students' disconnection from the course and result in learning outcomes not dependent on a students' ability to interact with their classmates. Further research on the utility of high centrality in the collaboration network of a course will benefit from a validated assessment of students' thermal physics understanding. This tool will help identify whether there are differences in learning gains between courses of well-connected versus more disconnected students.

Furthermore, future work on the analysis of student collaboration would benefit from qualitative interviews with students to investigate their beliefs and conceptions of group work. In particular, understanding students' thresholds for reporting helping or being helped by other students, the range of interactions that students have with their collaborators, and whether or not students find collaboration to be helpful will provide deeper insight into our results. Students' qualitative responses may shed light on what aspects of a course help to create a correlation between collaboration and performance. Furthermore, more data collection on in-person courses at CU Boulder would establish whether collaboration is associated with performance during regular instruction, or whether the environment at CU Boulder generally tends to have a weaker connection between student collaboration and performance.

\section*{Acknowledgements}

This work was funded in part by the CU Physics Department.  This material is based upon work supported by the National Science Foundation under Grants CCF-1839232, DMR-2002980, DGE-2125899, and PHY-2210566. We would like to thank the students who participated in the study and the members of PER@C and Carr Theoretical Physics groups for their general feedback. In particular, we extend a special thanks to Mina Fasihi Dolatabadi for helpful discussions on network analysis, Michael Vignal for assistance with the data analysis processes, and Giaco Corsiglia for providing feedback during our initial round of editing.

\bibliography{bibfile}

\begin{thebibliography}{21}%
\makeatletter
\providecommand \@ifxundefined [1]{%
 \@ifx{#1\undefined}
}%
\providecommand \@ifnum [1]{%
 \ifnum #1\expandafter \@firstoftwo
 \else \expandafter \@secondoftwo
 \fi
}%
\providecommand \@ifx [1]{%
 \ifx #1\expandafter \@firstoftwo
 \else \expandafter \@secondoftwo
 \fi
}%
\providecommand \natexlab [1]{#1}%
\providecommand \enquote  [1]{``#1''}%
\providecommand \bibnamefont  [1]{#1}%
\providecommand \bibfnamefont [1]{#1}%
\providecommand \citenamefont [1]{#1}%
\providecommand \href@noop [0]{\@secondoftwo}%
\providecommand \href [0]{\begingroup \@sanitize@url \@href}%
\providecommand \@href[1]{\@@startlink{#1}\@@href}%
\providecommand \@@href[1]{\endgroup#1\@@endlink}%
\providecommand \@sanitize@url [0]{\catcode `\\12\catcode `\$12\catcode
  `\&12\catcode `\#12\catcode `\^12\catcode `\_12\catcode `\%12\relax}%
\providecommand \@@startlink[1]{}%
\providecommand \@@endlink[0]{}%
\providecommand \url  [0]{\begingroup\@sanitize@url \@url }%
\providecommand \@url [1]{\endgroup\@href {#1}{\urlprefix }}%
\providecommand \urlprefix  [0]{URL }%
\providecommand \Eprint [0]{\href }%
\providecommand \doibase [0]{https://doi.org/}%
\providecommand \selectlanguage [0]{\@gobble}%
\providecommand \bibinfo  [0]{\@secondoftwo}%
\providecommand \bibfield  [0]{\@secondoftwo}%
\providecommand \translation [1]{[#1]}%
\providecommand \BibitemOpen [0]{}%
\providecommand \bibitemStop [0]{}%
\providecommand \bibitemNoStop [0]{.\EOS\space}%
\providecommand \EOS [0]{\spacefactor3000\relax}%
\providecommand \BibitemShut  [1]{\csname bibitem#1\endcsname}%
\let\auto@bib@innerbib\@empty
\bibitem [{\citenamefont {Hake}(1998)}]{Hake1998}%
  \BibitemOpen
  \bibfield  {author} {\bibinfo {author} {\bibfnamefont {R.~R.}\ \bibnamefont
  {Hake}},\ }\bibfield  {title} {\bibinfo {title} {Interactive-engagement
  versus traditional methods: A six-thousand-student survey of mechanics test
  data for introductory physics courses},\ }\href
  {https://doi.org/10.1119/1.18809} {\bibfield  {journal} {\bibinfo  {journal}
  {American Journal of Physics}\ }\textbf {\bibinfo {volume} {66}},\ \bibinfo
  {pages} {64} (\bibinfo {year} {1998})},\ \Eprint
  {https://arxiv.org/abs/https://doi.org/10.1119/1.18809}
  {https://doi.org/10.1119/1.18809} \BibitemShut {NoStop}%
\bibitem [{\citenamefont {Crouch}\ and\ \citenamefont
  {Mazur}(2001)}]{Mazur2001}%
  \BibitemOpen
  \bibfield  {author} {\bibinfo {author} {\bibfnamefont {C.~H.}\ \bibnamefont
  {Crouch}}\ and\ \bibinfo {author} {\bibfnamefont {E.}~\bibnamefont {Mazur}},\
  }\bibfield  {title} {\bibinfo {title} {Peer instruction: Ten years of
  experience and results},\ }\href {https://doi.org/10.1119/1.1374249}
  {\bibfield  {journal} {\bibinfo  {journal} {American Journal of Physics}\
  }\textbf {\bibinfo {volume} {69}},\ \bibinfo {pages} {970} (\bibinfo {year}
  {2001})},\ \Eprint {https://arxiv.org/abs/https://doi.org/10.1119/1.1374249}
  {https://doi.org/10.1119/1.1374249} \BibitemShut {NoStop}%
\bibitem [{\citenamefont {Anderman}\ and\ \citenamefont
  {Freeman}(2004)}]{MotivatingStudents2004}%
  \BibitemOpen
  \bibfield  {author} {\bibinfo {author} {\bibfnamefont {L.~H.}\ \bibnamefont
  {Anderman}}\ and\ \bibinfo {author} {\bibfnamefont {T.~M.}\ \bibnamefont
  {Freeman}},\ }\href@noop {} {\emph {\bibinfo {title} {Motivating Students,
  Improving Schools: The Legacy of Carol Midgley}}},\ edited by\ \bibinfo
  {editor} {\bibfnamefont {P.~R.}\ \bibnamefont {Pintrich}}\ and\ \bibinfo
  {editor} {\bibfnamefont {M.~L.}\ \bibnamefont {Maehr}}\ (\bibinfo
  {publisher} {Elsevier JAI},\ \bibinfo {address} {Amsterdam; London},\
  \bibinfo {year} {2004})\BibitemShut {NoStop}%
\bibitem [{\citenamefont {Fischer}(2007)}]{Fischer2007}%
  \BibitemOpen
  \bibfield  {author} {\bibinfo {author} {\bibfnamefont {M.~J.}\ \bibnamefont
  {Fischer}},\ }\bibfield  {title} {\bibinfo {title} {Settling into campus
  life: Differences by race/ethnicity in college involvement and outcomes},\
  }\href@noop {} {\bibfield  {journal} {\bibinfo  {journal} {Journal of Higher
  Education}\ }\textbf {\bibinfo {volume} {78}},\ \bibinfo {pages} {125}
  (\bibinfo {year} {2007})}\BibitemShut {NoStop}%
\bibitem [{\citenamefont {Zwolak}\ \emph {et~al.}(2017)\citenamefont {Zwolak},
  \citenamefont {Dou}, \citenamefont {Williams},\ and\ \citenamefont
  {Brewe}}]{Zwolak2017}%
  \BibitemOpen
  \bibfield  {author} {\bibinfo {author} {\bibfnamefont {J.~P.}\ \bibnamefont
  {Zwolak}}, \bibinfo {author} {\bibfnamefont {R.}~\bibnamefont {Dou}},
  \bibinfo {author} {\bibfnamefont {E.~A.}\ \bibnamefont {Williams}},\ and\
  \bibinfo {author} {\bibfnamefont {E.}~\bibnamefont {Brewe}},\ }\bibfield
  {title} {\bibinfo {title} {Students' network integration as a predictor of
  persistence in introductory physics courses},\ }\href
  {https://doi.org/10.1103/PhysRevPhysEducRes.13.010113} {\bibfield  {journal}
  {\bibinfo  {journal} {Phys. Rev. Phys. Educ. Res.}\ }\textbf {\bibinfo
  {volume} {13}},\ \bibinfo {pages} {010113} (\bibinfo {year}
  {2017})}\BibitemShut {NoStop}%
\bibitem [{\citenamefont {Zwolak}\ \emph {et~al.}(2018)\citenamefont {Zwolak},
  \citenamefont {Zwolak},\ and\ \citenamefont {Brewe}}]{Zwolak2018}%
  \BibitemOpen
  \bibfield  {author} {\bibinfo {author} {\bibfnamefont {J.~P.}\ \bibnamefont
  {Zwolak}}, \bibinfo {author} {\bibfnamefont {M.}~\bibnamefont {Zwolak}},\
  and\ \bibinfo {author} {\bibfnamefont {E.}~\bibnamefont {Brewe}},\ }\bibfield
   {title} {\bibinfo {title} {Educational commitment and social networking: The
  power of informal networks},\ }\href
  {https://doi.org/10.1103/PhysRevPhysEducRes.14.010131} {\bibfield  {journal}
  {\bibinfo  {journal} {Phys. Rev. Phys. Educ. Res.}\ }\textbf {\bibinfo
  {volume} {14}},\ \bibinfo {pages} {010131} (\bibinfo {year}
  {2018})}\BibitemShut {NoStop}%
\bibitem [{\citenamefont {Dou}\ \emph {et~al.}(2016)\citenamefont {Dou},
  \citenamefont {Brewe}, \citenamefont {Zwolak}, \citenamefont {Potvin},
  \citenamefont {Williams},\ and\ \citenamefont {Kramer}}]{Dou2016}%
  \BibitemOpen
  \bibfield  {author} {\bibinfo {author} {\bibfnamefont {R.}~\bibnamefont
  {Dou}}, \bibinfo {author} {\bibfnamefont {E.}~\bibnamefont {Brewe}}, \bibinfo
  {author} {\bibfnamefont {J.~P.}\ \bibnamefont {Zwolak}}, \bibinfo {author}
  {\bibfnamefont {G.}~\bibnamefont {Potvin}}, \bibinfo {author} {\bibfnamefont
  {E.~A.}\ \bibnamefont {Williams}},\ and\ \bibinfo {author} {\bibfnamefont
  {L.~H.}\ \bibnamefont {Kramer}},\ }\bibfield  {title} {\bibinfo {title}
  {Beyond performance metrics: Examining a decrease in students' physics
  self-efficacy through a social networks lens},\ }\href
  {https://doi.org/10.1103/PhysRevPhysEducRes.12.020124} {\bibfield  {journal}
  {\bibinfo  {journal} {Phys. Rev. Phys. Educ. Res.}\ }\textbf {\bibinfo
  {volume} {12}},\ \bibinfo {pages} {020124} (\bibinfo {year}
  {2016})}\BibitemShut {NoStop}%
\bibitem [{\citenamefont {Brewe}\ \emph {et~al.}(2012)\citenamefont {Brewe},
  \citenamefont {Kramer},\ and\ \citenamefont {Sawtelle}}]{Brewe2012}%
  \BibitemOpen
  \bibfield  {author} {\bibinfo {author} {\bibfnamefont {E.}~\bibnamefont
  {Brewe}}, \bibinfo {author} {\bibfnamefont {L.}~\bibnamefont {Kramer}},\ and\
  \bibinfo {author} {\bibfnamefont {V.}~\bibnamefont {Sawtelle}},\ }\bibfield
  {title} {\bibinfo {title} {Investigating student communities with network
  analysis of interactions in a physics learning center},\ }\href
  {https://doi.org/10.1103/PhysRevSTPER.8.010101} {\bibfield  {journal}
  {\bibinfo  {journal} {Phys. Rev. ST Phys. Educ. Res.}\ }\textbf {\bibinfo
  {volume} {8}},\ \bibinfo {pages} {010101} (\bibinfo {year}
  {2012})}\BibitemShut {NoStop}%
\bibitem [{\citenamefont {Brewe}\ \emph {et~al.}(2016)\citenamefont {Brewe},
  \citenamefont {Bruun},\ and\ \citenamefont {Bearden}}]{Brewe2016}%
  \BibitemOpen
  \bibfield  {author} {\bibinfo {author} {\bibfnamefont {E.}~\bibnamefont
  {Brewe}}, \bibinfo {author} {\bibfnamefont {J.}~\bibnamefont {Bruun}},\ and\
  \bibinfo {author} {\bibfnamefont {I.~G.}\ \bibnamefont {Bearden}},\
  }\bibfield  {title} {\bibinfo {title} {Using module analysis for multiple
  choice responses: A new method applied to force concept inventory data},\
  }\href {https://doi.org/10.1103/PhysRevPhysEducRes.12.020131} {\bibfield
  {journal} {\bibinfo  {journal} {Phys. Rev. Phys. Educ. Res.}\ }\textbf
  {\bibinfo {volume} {12}},\ \bibinfo {pages} {020131} (\bibinfo {year}
  {2016})}\BibitemShut {NoStop}%
\bibitem [{\citenamefont {Wells}\ \emph {et~al.}(2020)\citenamefont {Wells},
  \citenamefont {Henderson}, \citenamefont {Traxler}, \citenamefont {Miller},\
  and\ \citenamefont {Stewart}}]{Wells2020}%
  \BibitemOpen
  \bibfield  {author} {\bibinfo {author} {\bibfnamefont {J.}~\bibnamefont
  {Wells}}, \bibinfo {author} {\bibfnamefont {R.}~\bibnamefont {Henderson}},
  \bibinfo {author} {\bibfnamefont {A.}~\bibnamefont {Traxler}}, \bibinfo
  {author} {\bibfnamefont {P.}~\bibnamefont {Miller}},\ and\ \bibinfo {author}
  {\bibfnamefont {J.}~\bibnamefont {Stewart}},\ }\bibfield  {title} {\bibinfo
  {title} {Exploring the structure of misconceptions in the force and motion
  conceptual evaluation with modified module analysis},\ }\href
  {https://doi.org/10.1103/PhysRevPhysEducRes.16.010121} {\bibfield  {journal}
  {\bibinfo  {journal} {Phys. Rev. Phys. Educ. Res.}\ }\textbf {\bibinfo
  {volume} {16}},\ \bibinfo {pages} {010121} (\bibinfo {year}
  {2020})}\BibitemShut {NoStop}%
\bibitem [{\citenamefont {Wheatley}\ \emph {et~al.}(2021)\citenamefont
  {Wheatley}, \citenamefont {Wells}, \citenamefont {Henderson},\ and\
  \citenamefont {Stewart}}]{Wheatley2021}%
  \BibitemOpen
  \bibfield  {author} {\bibinfo {author} {\bibfnamefont {C.}~\bibnamefont
  {Wheatley}}, \bibinfo {author} {\bibfnamefont {J.}~\bibnamefont {Wells}},
  \bibinfo {author} {\bibfnamefont {R.}~\bibnamefont {Henderson}},\ and\
  \bibinfo {author} {\bibfnamefont {J.}~\bibnamefont {Stewart}},\ }\bibfield
  {title} {\bibinfo {title} {Applying module analysis to the conceptual survey
  of electricity and magnetism},\ }\href
  {https://doi.org/10.1103/PhysRevPhysEducRes.17.010102} {\bibfield  {journal}
  {\bibinfo  {journal} {Phys. Rev. Phys. Educ. Res.}\ }\textbf {\bibinfo
  {volume} {17}},\ \bibinfo {pages} {010102} (\bibinfo {year}
  {2021})}\BibitemShut {NoStop}%
\bibitem [{\citenamefont {Vargas}\ \emph {et~al.}(2018)\citenamefont {Vargas},
  \citenamefont {Bridgeman}, \citenamefont {Schmidt}, \citenamefont {Kohl},
  \citenamefont {Wilcox},\ and\ \citenamefont {Carr}}]{Vargas2018}%
  \BibitemOpen
  \bibfield  {author} {\bibinfo {author} {\bibfnamefont {D.~L.}\ \bibnamefont
  {Vargas}}, \bibinfo {author} {\bibfnamefont {A.~M.}\ \bibnamefont
  {Bridgeman}}, \bibinfo {author} {\bibfnamefont {D.~R.}\ \bibnamefont
  {Schmidt}}, \bibinfo {author} {\bibfnamefont {P.~B.}\ \bibnamefont {Kohl}},
  \bibinfo {author} {\bibfnamefont {B.~R.}\ \bibnamefont {Wilcox}},\ and\
  \bibinfo {author} {\bibfnamefont {L.~D.}\ \bibnamefont {Carr}},\ }\bibfield
  {title} {\bibinfo {title} {Correlation between student collaboration network
  centrality and academic performance},\ }\href
  {https://doi.org/10.1103/PhysRevPhysEducRes.14.020112} {\bibfield  {journal}
  {\bibinfo  {journal} {Phys. Rev. Phys. Educ. Res.}\ }\textbf {\bibinfo
  {volume} {14}},\ \bibinfo {pages} {020112} (\bibinfo {year}
  {2018})}\BibitemShut {NoStop}%
\bibitem [{\citenamefont {Williams}\ \emph {et~al.}(2019)\citenamefont
  {Williams}, \citenamefont {Zwolak}, \citenamefont {Dou},\ and\ \citenamefont
  {Brewe}}]{Williams2019}%
  \BibitemOpen
  \bibfield  {author} {\bibinfo {author} {\bibfnamefont {E.~A.}\ \bibnamefont
  {Williams}}, \bibinfo {author} {\bibfnamefont {J.~P.}\ \bibnamefont
  {Zwolak}}, \bibinfo {author} {\bibfnamefont {R.}~\bibnamefont {Dou}},\ and\
  \bibinfo {author} {\bibfnamefont {E.}~\bibnamefont {Brewe}},\ }\bibfield
  {title} {\bibinfo {title} {Linking engagement and performance: The social
  network analysis perspective},\ }\href
  {https://doi.org/10.1103/PhysRevPhysEducRes.15.020150} {\bibfield  {journal}
  {\bibinfo  {journal} {Phys. Rev. Phys. Educ. Res.}\ }\textbf {\bibinfo
  {volume} {15}},\ \bibinfo {pages} {020150} (\bibinfo {year}
  {2019})}\BibitemShut {NoStop}%
\bibitem [{\citenamefont {Wasserman}\ and\ \citenamefont
  {Faust}(1994)}]{WassermanFaust}%
  \BibitemOpen
  \bibfield  {author} {\bibinfo {author} {\bibfnamefont {S.}~\bibnamefont
  {Wasserman}}\ and\ \bibinfo {author} {\bibfnamefont {K.}~\bibnamefont
  {Faust}},\ }\href {https://doi.org/10.1017/CBO9780511815478} {\emph {\bibinfo
  {title} {Social network analysis: Methods and applications}}}\ (\bibinfo
  {publisher} {Cambridge University Press},\ \bibinfo {year}
  {1994})\BibitemShut {NoStop}%
\bibitem [{\citenamefont {Brandes}(2008)}]{Brandes2008}%
  \BibitemOpen
  \bibfield  {author} {\bibinfo {author} {\bibfnamefont {U.}~\bibnamefont
  {Brandes}},\ }\bibfield  {title} {\bibinfo {title} {On variants of
  shortest-path betweenness centrality and their generic computation},\ }\href
  {https://doi.org/https://doi.org/10.1016/j.socnet.2007.11.001} {\bibfield
  {journal} {\bibinfo  {journal} {Social Networks}\ }\textbf {\bibinfo {volume}
  {30}},\ \bibinfo {pages} {136} (\bibinfo {year} {2008})}\BibitemShut
  {NoStop}%
\bibitem [{\citenamefont {Newman}(2018)}]{Newman_Networks}%
  \BibitemOpen
  \bibfield  {author} {\bibinfo {author} {\bibfnamefont {M.}~\bibnamefont
  {Newman}},\ }\href@noop {} {\emph {\bibinfo {title} {Networks}}}\ (\bibinfo
  {publisher} {Oxford University Press, New York},\ \bibinfo {year}
  {2018})\BibitemShut {NoStop}%
\bibitem [{\citenamefont {Latora}\ \emph {et~al.}(2017)\citenamefont {Latora},
  \citenamefont {Nicosia},\ and\ \citenamefont {Russo}}]{Cmplx_Nets}%
  \BibitemOpen
  \bibfield  {author} {\bibinfo {author} {\bibfnamefont {V.}~\bibnamefont
  {Latora}}, \bibinfo {author} {\bibfnamefont {V.}~\bibnamefont {Nicosia}},\
  and\ \bibinfo {author} {\bibfnamefont {G.}~\bibnamefont {Russo}},\
  }\href@noop {} {\emph {\bibinfo {title} {Complex Networks Principles, Methods
  and Applications}}}\ (\bibinfo  {publisher} {Cambridge University Press},\
  \bibinfo {year} {2017})\BibitemShut {NoStop}%
\bibitem [{\citenamefont {Squartini}\ \emph {et~al.}(2013)\citenamefont
  {Squartini}, \citenamefont {Picciolo}, \citenamefont {Ruzzenenti},\ and\
  \citenamefont {Garlaschelli}}]{Squartini2013}%
  \BibitemOpen
  \bibfield  {author} {\bibinfo {author} {\bibfnamefont {T.}~\bibnamefont
  {Squartini}}, \bibinfo {author} {\bibfnamefont {F.}~\bibnamefont {Picciolo}},
  \bibinfo {author} {\bibfnamefont {F.}~\bibnamefont {Ruzzenenti}},\ and\
  \bibinfo {author} {\bibfnamefont {D.}~\bibnamefont {Garlaschelli}},\
  }\bibfield  {title} {\bibinfo {title} {Reciprocity of weighted networks},\
  }\href {https://doi.org/10.1038/srep02729} {\bibfield  {journal} {\bibinfo
  {journal} {Scientific Reports}\ }\textbf {\bibinfo {volume} {3}},\ \bibinfo
  {pages} {2729} (\bibinfo {year} {2013})}\BibitemShut {NoStop}%
\bibitem [{\citenamefont {Garlaschelli}\ \emph {et~al.}(2010)\citenamefont
  {Garlaschelli}, \citenamefont {Ruzzenenti},\ and\ \citenamefont
  {Basosi}}]{Garlaschelli2010}%
  \BibitemOpen
  \bibfield  {author} {\bibinfo {author} {\bibfnamefont {D.}~\bibnamefont
  {Garlaschelli}}, \bibinfo {author} {\bibfnamefont {F.}~\bibnamefont
  {Ruzzenenti}},\ and\ \bibinfo {author} {\bibfnamefont {R.}~\bibnamefont
  {Basosi}},\ }\bibfield  {title} {\bibinfo {title} {Complex networks and
  symmetry i: A review},\ }\href {https://doi.org/10.3390/sym2031683}
  {\bibfield  {journal} {\bibinfo  {journal} {Symmetry}\ }\textbf {\bibinfo
  {volume} {2}},\ \bibinfo {pages} {1683} (\bibinfo {year} {2010})}\BibitemShut
  {NoStop}%
\bibitem [{\citenamefont {Martin}\ and\ \citenamefont
  {Niemeyer}(2019)}]{martin_niemeyer_2019}%
  \BibitemOpen
  \bibfield  {author} {\bibinfo {author} {\bibfnamefont {C.}~\bibnamefont
  {Martin}}\ and\ \bibinfo {author} {\bibfnamefont {P.}~\bibnamefont
  {Niemeyer}},\ }\bibfield  {title} {\bibinfo {title} {Influence of measurement
  errors on networks: Estimating the robustness of centrality measures},\
  }\href {https://doi.org/10.1017/nws.2019.12} {\bibfield  {journal} {\bibinfo
  {journal} {Network Science}\ }\textbf {\bibinfo {volume} {7}},\ \bibinfo
  {pages} {180} (\bibinfo {year} {2019})}\BibitemShut {NoStop}%
\bibitem [{\citenamefont {Newman}(2008)}]{NewmanPhysicsToday}%
  \BibitemOpen
  \bibfield  {author} {\bibinfo {author} {\bibfnamefont {M.}~\bibnamefont
  {Newman}},\ }\bibfield  {title} {\bibinfo {title} {The physics of networks},\
  }\href {https://doi.org/10.1063/1.3027989} {\bibfield  {journal} {\bibinfo
  {journal} {Physics Today}\ }\textbf {\bibinfo {volume} {61}},\ \bibinfo
  {pages} {33} (\bibinfo {year} {2008})},\ \Eprint
  {https://arxiv.org/abs/https://doi.org/10.1063/1.3027989}
  {https://doi.org/10.1063/1.3027989} \BibitemShut {NoStop}%
\end{thebibliography}%

\end{document}